\documentclass[12pt]{article}
\usepackage{makeidx}
\makeindex

\newtheorem{Theorem}{Theorem}[section]

\newtheorem{Proposition}[Theorem]{Proposition}

\usepackage{latexsym}
\usepackage[T1]{fontenc}
\usepackage{amsmath}
\usepackage{amssymb}

\def \AO {{\cal A}({\cal O})}
\def \AO' {{\cal A}({\cal O}')}

\def \Pf {{\bf Proof.\,\,}}

\def \limR {\lim_{R \ra \infty}}

\def \be {\begin{equation}}
\def \ee {\end{equation}}
\def \ume {{\scriptstyle{\frac{1}{2}}}}
\def \ra {\rightarrow}

\def \eqq {\equiv}

\def \a {{\alpha}}
\def \b {{\beta}}
\def \g {{\gamma}}
\def \d {{\delta}}
\def \eps {{\varepsilon}}

\def \l {{\lambda}}
\def \La {{\Lambda}}
\def \s {{\sigma}}

\def \ph {{\varphi}}

\def \om {{\omega}}

\def \A {{\cal A}}

\def \D {{\cal D}}
\def \F {{\cal F}}
\def \G {{\cal G}}
\def \H {\mbox{${\cal H}$}}

\def \K {{\cal K}}
\def \L {{\cal L}}

\def \O {{\cal O}}
\def \P {{\cal P}}
\def \S {{\cal S}}

\def \U {{\cal U}}

\def \Z {{\cal Z}}

\def \id {{\bf 1 }}

\def \Psio {{\Psi_0}}

\def \di {{\partial_i}}
\def \dj {{\partial_j}}

\def \do {{\partial_0}}

\def \dmu {{\partial_\mu}}

\def \dt {{\partial_t}}

\def \d^nu {{\partial^\nu}}
\def \d^la {{\partial^\lambda}}
\def \d^o {{\partial^0}}

\def \abf {{\bf a}}

\def \hbf {{\bf h}}

\def \x {{\bf x}}
\def \y {{\bf y}}

\def \Abf {{\bf A}}

\def \Ebf {{\bf E}}

\def \Rbf {{\bf R}}

\def \nablabf {\mbox{\boldmath $\nabla$}}

\def\doppio#1{{\rm I}\kern-.1667em{\rm #1}}

\def\Q{\text{Q}\kern-.52em
    \text{\vrule height1.5ex width.5pt depth0pt}\kern.45em}

\def\dZ{{\mathchoice {\hbox{$\Ss\textstyle Z\kern-0.4em Z$}}
{\hbox{$\Ss\textstyle Z\kern-0.4em Z$}} {\hbox{$\Ss\scriptstyle
Z\kern-0.25em Z$}} {\hbox{$\Ss\scriptscriptstyle Z\kern-0.2em
Z$}}}}

\def\dC{{\mathchoice{\hbox{$\rm\textstyle\text{\kern.35em\vrule
   height1.5ex width.5pt depth0pt\kern-.35em C}$}}
{\hbox{$\rm\textstyle\text{\kern.35em\vrule
   height1.5ex width.5pt depth0pt\kern-.35em C}$}}
{\hbox{$\rm\scriptstyle\text{\kern.35em\vrule
   height1.5ex width.3pt depth0pt\kern-.35em C}$}}
{\hbox{$\rm\scriptscriptstyle\text{\kern.35em\vrule
   height1.5ex width.2pt depth0pt\kern-.35em C}$}}}}

\makeatletter

\@addtoreset{equation}{section}

\relax \@setckpt{pref0}{ \setcounter{page}{1}
\setcounter{equation}{0} \setcounter{enumi}{0}
\setcounter{enumii}{0} \setcounter{enumiii}{0}
\setcounter{enumiv}{0} \setcounter{footnote}{0}
\setcounter{mpfootnote}{0}
\setcounter{subsection}{0} \setcounter{subsubsection}{0}
\setcounter{paragraph}{0} \setcounter{subparagraph}{0}
\setcounter{figure}{0} \setcounter{table}{0}
\setcounter{Theorem}{0} }

\makeatletter

\begin{document}

\title{ Chiral symmetry breaking and $\theta$ vacuum structure in QCD}
\author{G. Morchio \\Dipartimento di Fisica, Universit\`a di Pisa,
\\and  INFN, Sezione di Pisa, Pisa, Italy \and F. Strocchi \\
Scuola Normale Superiore, Pisa, Italy \\ and INFN, Sezione di
Pisa}

\fussy

\date{}

\maketitle
\begin{abstract}The solution of the axial $U(1)$ problem, the role
of the  topology of the gauge group in forcing the breaking of
axial symmetry in any irreducible representation of the observable
algebra and the $\theta$ vacua structure are revisited in the
temporal gauge with attention to the mathematical consistency of
the derivations. Both realizations with  strong and weak Gauss law
are discussed; the control of the general mechanisms and
structures is obtained on the basis of the localization of the
(large) gauge transformations and the local generation of the
chiral symmetry. The Schwinger model in the temporal gauge exactly
reproduces the general results.

\end{abstract}

\date{}

\vspace{30mm}


\noindent \vspace{5mm}PACS: 11.30.Rd; 11.30.Qc; 12.38.-t; 11.40.Ha

\makeatletter
\newpage
\section{Introduction}
The solution of the $U(1)$ problem by the discovery of the
$\theta$ vacuum structure and its strong relation with chiral
symmetry breaking has been one of the cornerstones in the
theoretical analysis of the standard model of elementary
particles.~\cite{GH}~\cite{JR}~\cite{CDG}

The standard (and historically the first) arguments in favor of a
non-trivial role of the topology relies on semiclassical
approximations, in terms of boundary conditions of the classical
field configurations and their effect in the (euclidean)
functional integral. These results have been excellently reviewed
by Coleman,~\cite{C}~\cite{W} who also pointed out the limitation
of such arguments, since smooth field configurations, in
particular those of finite action, have zero functional measure.
The relevance of the classification of the (smooth) field
configurations in terms of their pure gauge behaviour at space
infinity, and of the corresponding winding number, is therefore a
non-trivial mathematical problem, especially in the infinite
volume limit.

 A decisive step in the
direction of a mathematical control of the mechanism of $\theta$
vacuum structure was taken by Jackiw, who emphasized the role of
the topology of the gauge group, without involving the
classification  of the gauge field configurations and thus
avoiding the problem of the semiclassical approximation of the
functional integral.~\cite{J}

Jackiw's analysis is done in the temporal gauge and the very
peculiar features of such a gauge raise some structural
mathematical problems, with a non-trivial impact on the general
argument. The aim of this note is to present an analysis which
does not suffer of mathematical inconsistencies and, in a certain
sense, provides a mathematical glossary to Jackiw's strategy.

The paper is organized according to the following pattern.

As in the abelian case, the invariance of the vacuum under the
gauge transformations generated by the Gauss operator implies a
{\em non-regular representation of the field algebra} (Section 2).

In Section 3, the {\em non-trivial topology} of the group $\G$ of
time independent gauge transformations is classified in terms {\em
of group valued gauge functions $\U(\x)$ of compact
support}.\goodbreak

This significantly simplifies the discussion with respect to the
conventional analysis,~\cite{J}~\cite{C} where the gauge functions
are only required to have a limit, for $|\x| \ra \infty$,
independent of the direction (with the need of requiring for the
analysis a faster than $1/r$ decay of the gauge vector fields
$A_i^a(\x)$). Furthermore, the exponentials of the topological
charge (in bounded regions), at the basis of the standard
analysis, are shown to have vanishing matrix elements between
vectors satisfying the Gauss law constraint and therefore a
non-trivial impact of the large gauge transformations on the
structure of the physical states requires further ingredients.

In Section 4, it is argued that a crucial role for disclosing the
physical relevance of the non-trivial topology of the gauge group
is played by the fermions and the associated {\em chiral
symmetry}.

Following Bardeen,~\cite{B} we show that, contrary to statements
appeared in the literature, the presence of the chiral anomaly
does not prevent the  chiral symmetry from being { a well defined
time independent group of automorphisms of the field algebra} and
of its gauge invariant (observable) subalgebra, generated by the
operators $V_R^5(\l)$, $\l \in \Rbf$, formally the exponentials of
the charge density $J_0(f_R \a_R)$, $J_\mu^5$ denoting the
conserved gauge dependent axial current .

The {\em solution of the $U(1)$ problem} (i.e. the absence of
massless Goldstone bosons associated to chiral symmetry breaking)
is provided by the failure of a crucial hypotheses of the
Goldstone theorem, namely the impossibility of writing the
symmetry breaking Ward identities, since the vacuum expectations
$< J_0^5(f_R\,\a_R) \,\bar{\psi}\,\psi >_0$ do not exist, as a
consequence of the non-regularity of the local implementers of the
chiral symmetry, $V_R^5(\l)$.

The interplay between the {\em topology of the gauge group} and
the chiral transformations which {\em forces chiral symmetry
breaking} (in any irreducible, or even factorial,  representation
of the observable algebra) and gives rise to a non-trivial vacuum
structure is clearly displayed under general assumptions. By
assuming that the localized large gauge transformations may be
implemented by unitary operators which are ``functions'' of fields
with the same localization,  we show that the non-trivial topology
of the gauge group reflects in a {\em non-trivial center of the
algebra of observables, which is not left pointwise invariant
under the chiral transformations} (Section 5).

The standard  labeling of the irreducible (or even factorial)
representations of the observable algebra by an angle $\theta \in
[0, \pi)$, ($\theta$ {\em sectors}) is obtained by analyzing the
reducible representation of the field algebra defined by a chiral
invariant vacuum and by a spectral decomposition over the center
of the observable algebra.

In Section 6, we consider a realization of the temporal gauge in
which only a weak Gauss invariance of the vacuum is required, as
discussed in the abelian case.~\cite{LMS} As a consequence of the
lack of Gauss invariance of the vacuum, the conserved axial
current $J_\mu^5$ can be represented by a well defined field
operator and Bardeen analysis directly applies.

 Under the same assumption of implementation of the large
gauge transformations by local operators, we show that, if the
chiral symmetry is unitarily implemented,
  the vacuum  defines a
reducible representation $\pi^0$ of the observable algebra; chiral
symmetry is broken in each irreducibile component. Independently
of the possibility of extending the $\theta$ vacua to non-positive
weakly gauge invariant functionals on the field algebra, the
chiral current $J_\mu^5$ does not exist in the physical
representation space of the observable algebra, in particular in a
$\theta$ sector.

It is worthwhile to stress that, contrary to statements appeared
in the literature, the derivation of the physical consequences of
the topology of $\G$ crucially  relies on the presence of fermions
and their chiral transformations. The essential point, which
distinguishes the abelian case from the non-abelian one, beyond
the existence of the chiral anomaly,  is the {\em existence of a
center of the local observable algebras, following from the
topology of $\G$, which is not pointwise invariant under the
chiral transformations}.

All the general results are exactly reproduced by the Schwinger
model in the temporal gauge, analyzed in Section 7, both in the
positive non-regular realization with a Gauss invariant vacuum and
in the indefinite regular (quasi free) realization, with the weak
form of the Gauss law constraint.

%%%%%%%%%%%%%%%%%%%%%%%%%%%%%%%%%%%%%%%%%%%%%%%%%%%%%%%%%%%%
%%%%%%%%%%%%%%%%%%%%%%%%%%%%%%%%%%%%%%%%%%%%%%%%%%%%%%%%%%%%

%%%%%%%%%%%%%%%%%%%%%%%%%%%%%%%%%%%%%%%%%%%%%%%%%%%%%%%%%%%%
%%%%%%%%%%%%%%%%%%%%%%%%%%%%%%%%%%%%%%%%%%%%%%%%%%%%%%%

\newpage
\sloppy
\section
{Temporal gauge in QCD and Gauss law}\fussy
%\addtocontents{toc}{Temporal gauge in QCD. Vacuum structure}
\index{QCD} \index{temporal gauge in QCD} \index{QDC!vacuum
structure} For the discussion of the non-perturbative aspects of
QCD, in particular the $\theta$ vacuum structure, its relation
with the topology of the gauge field configurations and its role
in chiral symmetry breaking, the temporal gauge has proved to be
particularly convenient.

However, as noted before for the abelian (QED) case,~\cite{LMS}
 the conflict between the Gauss law constraint and
canonical quantization raises problems of mathematical
consistency, which, as we shall see, affect the derivation of the
general structures leading to significant physical implications.

\def \uqua {{\scriptstyle{\frac{1}{4}}}}
\def \Ebf {{\bf E}}
\def \Bbf {{\bf B}}
\def \Abf {{\bf A}}
\def \Dbf {{\bf D}}
\def \Gbf {{ \bf G}}
\def \DE {{ \bf D \cdot E}}
\def \nablabf {\mbox{\boldmath $\nabla$}}
\def \d {\delta}
\vspace{2mm}
 For simplicity, we start by considering  the case with
only vector fields (no fermion or scalar field being present).
Then, at the classical level the QCD Lagrangean density reduces to
the Yang-Mills form \be{ \L =  - \uqua \sum_a
F_{\mu\,\nu}^a\,F^{\mu\,\nu\,\,a} = \ume \sum_{a}(\Ebf_a^2 -
\Bbf_a^2),}\ee where in the temporal gauge, defined by  $A^0_a =
0$, \be{\Ebf _a = - \dot{\Abf}_a,\,\,\,\,\, \,\,\,\,\Bbf_a =
\nablabf \times \Abf_a - \ume g f_{a b c}\,\Abf_b \times \Abf
_c,}\ee ($a$ is a color index and $f_{a b c}$ are the structure
constants of the Lie algebra of the color gauge group $\G$).

The corresponding equations of motion, obtained by variations with
respect to $\Abf_a$, are \be{ \dt\, \Ebf_a = \nablabf \times
\Bbf_a + g f_{a b c } \Abf_b \times \Bbf_c \eqq (\Dbf \times
\,\Bbf)_a,}\ee which imply \be{ \dt G_a  = 0,\,\,\,\,G_a \eqq
\nablabf \cdot \Ebf_a + g f_{a b c}\,\Abf_b \cdot \Ebf_c \eqq
(\Dbf \cdot \Ebf)_a.}\ee The operators $G_a$ are called the {\em
Gauss law operators}.

In the standard quantum version of the temporal gauge it is
assumed that the fields $\Abf_a$ and their powers can be defined
and quantization is given by the canonical commutation
relations.\goodbreak

In particular one has the following commutation relations \be{-i\,
[\, \DE_a(\x, t), \,\Abf_b(\y, t)\,] = \d_{a b}
\,\mbox{\boldmath$\nabla$} \d(\x - \y) + g f_{a b c} \Abf_c(\x,
t)\, \d(\x-\y).}\ee $$-i\, [\, \DE_a(\x, t), \,\Ebf_b(\y, t) \,] =
g f_{a b c}\, \Ebf_c(\x,t)\,\d(\x- \y).$$ They state that the
Gauss operators generate  the infinitesimal time independent gauge
transformations, $\d^\La$, with $\La^a(\x) \in \S(\Rbf^3)$ the
c-number gauge function \be{ \d^{\La} \Abf_a(x) = \nablabf \La(\x)
+ g f_{a b c } \Abf^b(x)\,\La^c(\x) ).}\ee Since the variables
$A_a^0$ are missing in the Lagrangean, one cannot exploit the
stationarity of the action with respect to them and therefore one
does not get the Gauss law $G_a = 0$. Actually, the Gauss law is
incompatible with eq.\,(2.5) and therefore with canonical
quantization and more crucially with the Gauss operator being the
generator of the  time independent gauge transformations,
eq.\,(2.6).

A proposed solution of this conflict, widely adopted in the
literature and in textbook discussions of the temporal gauge, is
to require  the Gauss law constraint as an operator equation on
the (subspace of) physical states and, in particular, on the
vacuum state. However, such a solution is not  mathematically
consistent. In fact, the vacuum expectation of eq.\,(2.6) gives
zero on the left hand side and non-zero on the right hand side.

It has been proposed~\cite{J}~\cite{BNS} to cope with this paradox
by admitting that the vacuum vector is not normalizable. In our
opinion, such a solution is not acceptable, because it does not
yield a representation of a  field algebra containing both gauge
dependent and gauge independent fields.

A mathematically acceptable solution for the Gauss law constraint
is to adopt a Weyl quantization and admit non-regular
representations.

As a  preliminary step in this direction, we recall that from a
mathematical point of view the quantum fields are operator valued
distributions and a smearing with test functions (typically
infinitely differentiable and of compact support) is needed for
obtaining well defined Hilbert space operators. Therefore, we
consider the {\em field algebra} generated by the polynomials of
the smeared fields $A^i_a(f)$, $f \in \S(\Rbf^4)$;   we shall
assume that by a suitable  point splitting procedure one can
consider as field variables the powers of $\Abf_a(x)$ and its
derivatives, like e.g. the Gauss operator $G_a(g)$, the
``magnetic'' field $B^i_a(g)$, $ \,g \in \S(\Rbf^4)$, etc.

We then take the polynomials of such field variables as a {\em
local} (Borchers) {\em field algebra} $\F$, transforming
covariantly under the space time translations $\a_y$, $y \in
\Rbf^4$, $$\a_y(A^i_a(f)) = A^i_a(f_{y}), \,\,\,\,\,\,\,f_{y}(x) =
f(x - y).$$

\def \fbf {{\bf f}}
\def \gbf  {{\bf g }}
\def \hbf {{\bf h }}

 In order to simplify the bookkeeping of the indices, it is convenient to
introduce the following notations: $T^a$ denote the hermitean
representation matrices of the Lie algebra of the gauge group,
normalized  so that Tr $ \,T^a\,T^b =  \d_{a\,b}$, the field
$A^i(x) =  \sum_a A^i_a T^a$ are Lie algebra valued distributions
on Lie algebra valued test functions $\fbf^i(x) = \sum_a
f^i_a(x)\,T^a$, $f^i_a \in \S(\Rbf^4)$, $$\Abf(\fbf) \eqq \int d^4
x \mbox{Tr}\,(\Abf(x)\,\fbf(x)) = \int d^4 x \sum_{i,\,a}
A^i_a(x)\,f^i_a(x).$$ Unless otherwise stated, in the following
the sum over repeated space and gauge indices will be understood.

With this notation the time independent gauge transformations
$\a_\U$ are labeled by gauge group valued unitary $C^\infty$
functions $\U(\x)$, which may be taken to differ from the identity
only on a compact set $K_\U$, and \be{\a_\U(\Abf(\fbf)) = \Abf(\U
\fbf \U^{-1}) + \U \mbox{\boldmath $\partial $} \U^{-1}(\fbf),}\ee
$$\U \mbox{\boldmath $\partial $} \U^{-1}(\fbf) \eqq \int d^4 x
\,\, \mbox{Tr}\,(\sum_i \U(\x)
\partial^i  \U^{-1}(\x)\,{\bf f}^i(x)).$$
\def \prbf {{\mbox{\boldmath $\partial $}}}
\def \trs  {transformation\;}
\def \trss {transformations\;}

The (space-time) localization of $\U$ is given by the cylinder
$C_\U \eqq \K_\U \times \Rbf$, so that $\a_\U(\Abf(\fbf)) =
\Abf(\fbf)$ if supp\,$\fbf \cap C_\U = \emptyset$.

We denote by $\U^\l$, $\l \in \Rbf$,  the gauge functions
corresponding to
 one-parameter subgroups of the gauge
group. They can be written in the form $\U^\l(\x) = e^{i\l\,
g(\x)}$ with $ g = \sum_a g_a(\x) \,T^a$ a  Lie algebra valued
function, infinitely differentiable and of compact support ($g_a
\in \D(\Rbf^3)$). All gauge transformations of compact support in
a neighborhood of the identity, in the $C^\infty$ topology, are of
this form; they generate the Gauss subgroup $\G_0$ of the gauge
group $\G$.\goodbreak

 For discussing  the Weyl quantization one has to consider the
{\em exponential field algebra} $\F_W$ generated by the unitary
operators $W(\fbf)$, $\fbf^i = \sum_a f^i_a(x) \,T^a$, $f^i_a \in
\S(\Rbf^4)$, formally  the exponentials $e^{ i \Abf(\fbf)}$, and
by unitary operators $V(\U^\l)$, representing $\G_0$, formally the
exponentials of the Gauss operators, $$V(\U^\l) =
e^{i\,\l\,G(g)},\,\,\,\,\,G(g) = \sum_a G_a(g_a), \,\,\,\,\,g_a
\in \D(\Rbf^3),$$  transforming covariantly under space
translations. Their time independence formally follows if the
dynamics $\a_t$ is generated by local gauge invariant Hamiltonians
$H_R$, so that \be{ (d/dt) \a_t(V(\U^\l)) = i \limR
[\,H_R,\,V(\U^\l)\,] = 0.}\ee

\def \exU {{e^{i \U \scriptsize{\prbf}  \U^{-1}(\fbf)}}}
 The {\em observable field subalgebra} of $\F_W$ is characterized
 by its pointwise invariance
under gauge transformations.

A representation of $\F_W$ also defines a representation of $\F$
only if it is {\bf regular}, i.e. if (the representatives of) the
field exponentials $W(\l \fbf)$, $\l \in \Rbf$ define weakly
continuous one-parameter groups.

A  state $\om$ on the exponential field algebra $\F_W$, in
particular a vacuum state, is said to satisfy the {\bf Gauss law}
in exponential form, if $\omega(V(\U^\l)) = 1$, equivalently if
its representative vector $\Psi_\om$ in the  GNS Hilbert space
$\H_\om$ (defined by the expectations of $\F_W$ on $\omega$)
satisfies \be{V(\U^\l)\,\Psi_\om =
\Psi_\om,\,\,\,\,\,\,\,\forall\, \U^\l.}\ee
 Briefly, a vector
state $\Psi$ satisfying eq.\,(2.9)  is said to be {\bf Gauss
invariant}. An operator in $\H$ is  {\em Gauss invariant} if it
commutes with all the $V(\U^\l)$.

In the following we shall consider the realization of the temporal
gauge defined by a vacuum state $\om$ satisfying the  Gauss law.
As in other interesting quantum mechanical models, like the
electron in a periodic potential (Bloch electron), the quantum
particle on a circle, the Quantum Hall electron, etc., the
invariance of the ground state under a group of gauge
transformations implies that the corresponding representation of
the exponential  field algebra is not
regular.~\cite{AMS}\goodbreak

\begin{Proposition} A  vacuum state $\om$ on the
exponential field algebra $\F_W$, satisfying the Gauss law,
defines a non-regular representation of $\F_W$, since :
\be{\hspace{20mm} \om(W(\fbf^i) ) = 0,
\,\,\,\,\,\,\,\,\,\,\mbox{if}\,\,\,\,\,{\bf f}^i(x)  \neq 0,}\ee
$$\hspace{31mm}= 1,\, \,\,\,\,\,\,\,\,\,\mbox{if}\,\,\,\,\,\fbf =
0.$$ The fields $\Abf$, formally the generators of the $W(\fbf)$,
cannot be defined in the  GNS Hilbert space defined by the vacuum
expectations and in particular the two point function of the gauge
potential does not exist, only (the vacuum expectations of) the
exponential functions (and of course the gauge invariant
functions) of $\Abf$ can be  defined.

In the free case, i.e. for vanishing gauge coupling constant, the
exponential field algebra becomes a Weyl field algebra, generated
by the exponentials of $\Abf_a$ and of its conjugate momenta
$\Ebf_a$, and eqs.\,(2.10) uniquely determine its representation
as a {\bf non-regular Weyl quantization}.
\end{Proposition}
\Pf\,\,For each  $\fbf^i$ there is a one-parameter subgroup
$\U^\l$ such that $\U^\l \,\fbf^i\, \U^{-\l} = \fbf^i$, and $
\exp{i D_{\U^\l}(\fbf^i)} \eqq  \exp{(i\, \U^\l \prbf\,
\U^{-\l}(\fbf^i))} \neq 1$; therefore,   by eq.\,(2.9), one has
 $$ \om(W(\fbf^i)  =
\om(V(\U^\l) \, W(\fbf^i)  \,V(\U^\l)^*) = e^{i D_{\U^\l}(\fbf^i)
} \,\om(W(\fbf^i)), $$ and eqs.\,(2.10) follow.

\noindent Clearly, the one-parameter groups defined by $W(\fbf)$
cannot be weakly continuous and therefore the corresponding
generators, i.e. the fields $A^i_a(f)$ do not exist as  operators
in the GNS Hilbert space defined by the expectations of $\F_W$ on
$\om$. The free case can be worked out along the same lines as for
the abelian case.~\cite{LMS}

\vspace{1mm} The Hilbert space $\H$ of the representation of
$\F_W$ defined by the vacuum state $\om$, satisfying the Gauss
law, contains a subspace $\H'$ of Gauss invariant vectors
$$V(\U^\l)\, \Psi = \Psi, \,\,\,\,\,\forall \Psi \in
\H',\,\,\,\,\,\,\forall\, \U^\l.$$ We denote by $\pi_0$ the
representation of $\G_0$ in $\H$ given by the $V(\U^\l)$.

It is worthwhile to remark that  the local  operator $v(\U^\l)
\eqq V(\U^\l) - \id$  is non zero in $\H$, since $V(\U^\l)$
implements the time independent gauge transformations,
corresponding to the one-parameter subgroups, which are
non-trivial on $\F_W$. Thus, the  assumptions of the
Reeh-Schlieder theorem, according to which a local operator which
annihilates the vacuum must vanish, cannot be satisfied.

One can easily check that the crucial point in the proof of the
theorem fails, namely $\F_W(\O)\,\Psio$, where $\F_W(\O)$ is the
exponential field algebra localized in the region $\O$, is not
dense in $\H$. In fact,  for any $\O$ disjoint from  $C_{\U^\l}$,
by locality one has $$ (v(\U^\l) \,\F_W \Psio, \, \F_W(\O)\,\Psio)
= (\F_W \Psio, \, \F_W(\O)\,v(\U^\l)^*\,\Psio) = 0,$$ $$v(\U^\l)\,
\F_W \Psio \neq 0.$$ Actually, the Reeh-Schlieder theorem does not
apply because the relativistic spectral condition fails in $\H$.
In fact, the implementers $U(\abf)$ of the space translations are
not weakly continuous, since,  as in the proof of Proposition 2.1,
$$\om(W(\fbf) U(\abf) W(-\fbf))= \om(W(\fbf)\,W(-\fbf_\abf)) =$$
$$=e^{i D_{\U^\l}(\fbf - \fbf_a)}\,\om(W(\fbf)\,W(-\fbf_\abf)),$$
so that the right hand side vanishes if $\abf \neq 0$ and it is $=
1$, otherwise. The spectral condition for the Fourier transforms
of the matrix elements of $U(\abf, t) = U(\abf) \,U(t)$ is
violated if there is strong continuity in $t$, since it would
imply that, after smearing in time, the Fourier transform in
$\abf$ is finite measure and therefore continuity in $\abf$ of the
above matrix elements.

\vspace{1mm} It is worthwhile to remark that the one-parameter
groups $V(\U^\l)$ are not assumed to be weakly continuous in $\l$;
actually,  continuity cannot hold if the global gauge group is
simple and has rank at least  two (as in the case of color
$SU(3)$), since then one  obtains the vanishing of $\om(W(\fbf)
V(U_\l) W(-\fbf))$, for $\l \neq 0$, $f^i_c(x)  = \d_{c a}
f^i(x)$, $\U^\l \fbf^i \U^{-\l} = f^i(x) T^b$, $[\,T^a, \,T^b\, ]
= 0$, from the invariance of $\om$ under the subgroup generated by
$T^a$, as  in  the proof of Proposition 2.1. Thus, in this case
the Gauss law constraint can only be imposed in the exponential
form.

%%%%%%%%%%%%%%%%%%%%%%%%%%%%%%%%%%%%%%%%%%%%%%%%%%%%%
%%%%%%%%%%%%%%%%%%%%%%%%%%%%%%%%%%%%%%%%%%%%%%%%%%%%%%%%%%

%%%%%%%%%%%%%%%%%%%%%%%%%%%%%%%%%%%%%%%%%%%%%%%%%%%%%%%
%%%%%%%%%%%%%%%%%%%%%
%%%%%%%%%%%%%%%%%%%%%%%%%%%%%

%%%%%%%%%%%%%%%%%%%%%%%%%%%%%%%%%%%%%%%%%%%%%%%%%
%%%%%%%%%%%%%%%%%%%%%%%%%%%%%%%%%%%%%%%%%%%%%%%%%%%

\newpage

\section{Topology of the gauge group} By definition,
eq.\,(2.7), we consider localized
gauge functions; they obviously extend to the one point
compactification of $\Rbf^3$,  $\dot{\Rbf}^3$,  which is
isomorphic to the three-sphere $S^3$,  $$\U(\x): \dot{\Rbf}^3 \sim
S^3 \ra \G.$$

Such maps fall into disjoint homotopy classes labeled by
``winding'' numbers $n$ $$ n(\U) = (24 \pi^2)^{-1} \int d^3x\,
\eps^{i j k}\, \mbox{Tr} \,( \U_i(\x)\,\U_j(\x)\,\U_k(\x)) \eqq
\int d^3 x \,n_\U(\x),$$ where $ \U_i(\x) \eqq
\U(\x)^{-1}\,\di\,\U(\x)$.

The gauge transformations with $n \neq 0$ are called {\em large
gauge transformations}. Those with zero winding number are called
{\em small}; since they are contractible to the identity, they are
products of $\U(\x)$ which are close to the identity (in the
$C^\infty$ topology) and therefore are expressible as products of
$\U^\l$, i.e. they are elements of $\G_0$. Clearly, all elements
of $\G_0$ have zero winding number.

The following analysis may be also applied to gauge
transformations $\U(\x)$ which are only required to have a limit
for $|\x| \ra \infty$, since they are of compact support modulo
Gauss transformations $\U^\l$ with such a behaviour; only  the
existence of the corresponding Gauss operators
 $V(\U^\l)$ is required, with no additional implications.

In the above realization  of the temporal gauge, the small gauge
transformations  are implemented by the unitary operators
$V(\U^\l) \in \F_W$; the next question is the implementability of
the large gauge transformations and their distinction from the
small on the Gauss invariant states. In fact, the non-triviality
of the large gauge transformations on the physical space turns out
to be a rather subtle question as displayed by the following
Proposition.

\begin{Proposition} A local field operator invariant under Gauss gauge
transformations is also invariant under large gauge
transformations.

Any  vector $\Psi \in \H'$, in particular the vacuum vector,
defines a state, i.e. expectations, on $\F_W$ invariant also under
the large gauge  transformations.

\end{Proposition}
\Pf\,\, In fact, given a gauge function $\U_n(\x)$ and its space
translated by $\abf$,  $\U^a_n(\x) = \U_n(\x - \abf)$, the
combined gauge transformations $\a_{\U^a_n}^{-1}\,\a_{\U_n}$ and
$\a_{\U_n}\,\a_{\U^a_n}^{-1}$ have zero winding number and
therefore are small gauge transformations, say $\a_{\U_0}$ and
$\a_{\U'_0}$ respectively.

\noindent Then, for any local field operator $F$ invariant under
small gauge transformations one has $$\a_{\U_n}(F) = \a_{\U^a_n}
\,\a_{\U_0} (F) = \a_{\U^a_n}(F),$$ and for $|\abf|$ sufficiently
large, by locality $\a_{\U^a_n}(F) = F$.

 \noindent Quite generally, for any (local) operator $F \in
\F_W$ one has, for $|\abf|$ sufficiently large, $$ \a_{\U_n}(F) =
\a_{\U'_0} \a_{\U^a_n}(F) = \a_{\U'_0}(F);$$  by the Gauss
invariance of $\om$, this implies $\om(\a_{\U_n}(F)) = \om(F).$
%This also proves the last statement.

\vspace{2mm} By a standard argument, the invariance of the vacuum
under the large gauge transformations implies that the gauge
transformations are implemented by unitary operators; they can be
chosen  to represent the gauge group $\G$, to coincide the
$V(\U^\l)$, for all Gauss transformations, and    to transform
covariantly under space translations.

 Furthermore, if, as we assume, the dynamics is generated
by local gauge invariant Hamiltonians, the implementers commute
with the time translations. The implementers are unique, up to
phases,  if the field algebra is irreducible in $\H$; in this case
the implementers of the large gauge transformations are multiples
of the identity in $\H'$.

The above results indicate that the distinction between the small
and the large transformations at the level of physical states is
problematic and a crucial question is the implication, if any, of
the non-trivial topology of the gauge group on the physical
states, more generally on the representations of the observable
algebra.

 One of the standard (and
historically the first) arguments in favor of a non-trivial role
of the topology of the gauge transformations relies on
semiclassical approximations, in terms of boundary conditions of
the classical field configurations and their effect on the
(euclidean) functional integral. This approach has been
excellently reviewed by Coleman,~\cite{C} who also pointed out the
limitation of such arguments, since smooth field configurations,
in particular those of finite action, have zero functional
measure.

The relevance of the classification of the (smooth) field
configurations in terms of their pure gauge behaviour at space
infinity, and of the corresponding winding number, is therefore a
non-trivial mathematical problem, especially in the infinite
volume limit. The topological classification is done in finite
volume, but in the infinite volume limit a non-trivial instanton
density implies, in the dilute gas approximation, that the
functional measure is concentrated on configurations with
divergent topological number. The association of the non-trivial
topology with the existence of instanton solutions, which minimize
the classical action and as such do not have compact support, is
probably the reason why  localized  large gauge transformations
have not been considered in the literature.

%Quite generally, the construction of the $\theta $ vacua in terms
%of the $n$-vacua is inconsistent in infinite volume and in fact,
%by Proposition 2.2 all the translationally invariant $n$-states
%defined by  the vectors $T_n\,\Psio$, which are the analogs of the
%$n$-vacua, give all the same expectations of (the exponential
%field algebra) $\F_W$. In particular, the euclidean correlations
%functions, obtained  by working in a finite euclidean volume $V
%\times T$ with boundary conditions $A_\mu = \U_n
%\partial_\mu \U_n^{-1}$ on the space boundary $\partial V$, define,
%in the thermodynamical limit, translationally invariant
%states,  $\omega_n$, i.e. expectations $\omega_n(\F_W)$,
%invariant under large transformations and independent of $n$
%$$\omega_n(F) = \omega_n(V(\U_m) F V(\U_m)^{-1}) =
%\omega_{n+m}(F).$$

%The functional integral construction of the $\theta$ vacua changes
%drastically in the presence of fermions, since in this case the
%expectation of a product of fields with total chirality $2n$ has
%contributions only from functional integral configurations with
%topological number $n$, whereas without fermions only
%configurations with infinite topological number are relevant in
%the infinite volume.

As we shall see  below, the presence of fermions plays a crucial
role for the non-trivial effects  of the topology of the gauge
group.

%%%%%%%%%%%%%%%%%%%%%%%%%%%%%%%%%%%%%%%%%%%%%%%
%%%%%%%%%%%%%%%%%%%%%%%%%%%%%%%%%%%%%%%%%%%%%%%
\vspace{1mm} Another argument for the physical consequences of the
gauge group topology has been proposed in terms of the topological
charge.~\cite{J} The so-called {\em topological current}  is
formally defined by \be{ C^\mu(x) = - (16 \pi^2)^{-1} \eps^{\mu
\nu \rho \s} \mbox{Tr} (F_{\nu \rho}(x)\,A_\s(x) -
{{\scriptstyle{\frac{2}{3}}}} A_\nu(x)\,A_\rho(x)\,A_\s(x)).}\ee
$$\dmu C^\mu(x) =  - ( 16 \pi^2)^{-1} \,\mbox{Tr}\,^*F_{\mu
\nu}(x)\,F_{\mu \nu}(x) \eqq \P,$$ where $A_\mu = (0, A_i)$,
$^*F_{\mu \nu} \eqq \eps_{\mu \nu \rho \s} \,F^{\rho \s}$. In the
mathematical literature, for classical fields, $\P$ is called the
``Pontryagin density'' and $C_\mu$ the ``Chern-Simons secondary
characteristic class''.

At the classical level, one has~\cite{J} the following
transformation law of $C_0(x)$ under gauge transformations
$\a_\U$, defined by eqs.\,(2.7), \be{ \a_{\U}(C_0(x)) = C_0(x)
 -(8 \pi^2)^{-1} \di[ \eps^{i j k} \mbox{Tr}(\dj \U(\x)\, \U(\x)^{-1}
A_k(x))] + n_{\U}(\x).}\ee Therefore, at the classical level the
space integral of $C_0(\x,x_)$ is invariant under small
transformations, but it get shifted by $n$ under gauge
transformations with winding number $n$.\goodbreak

 For the quantum case one
meets non-trivial consistency problems. First of all the formal
expression in the right hand side of eq.\,(3.3) requires a point
splitting regularization. It is reasonable to assume that this can
be done by keeping the transformation properties of the formal
expression under large gauge transformations,
eq.\,(3.4).\goodbreak

The next problem is the  space integral of $C_0(x)$. The space
integrals of charge densities, even for conserved currents, are
known to diverge and  suitable regularizations are needed,
including a time smearing (see eq.\,(3.5) below). In the case of
conserved currents, under some general conditions one may obtain
the convergence of a suitably regularized  integral of the charge
density,   in matrix elements on states with suitable
localization;~\cite{SS}~\cite{M}~\cite{R}~\cite{MS} but in the
general case the problem seems to be open.

Actually, in the quantum  temporal gauge an even more serious
problem arises by the gauge dependence of $C_\mu(x)$. By
Proposition 2.1,  the regularized space integral of $C_0(x)$
\be{C_0(f_R \a_R )\eqq \int d^4 x
\,f_R(\x)\,\a_R(x_0)\,C_0(x),}\ee where $f_R(\x) = f(|\x|/R)$,
$f(x) = 1$, for $|x| \leq 1$, $= 0$, for $|x| \geq 1 + \eps$,
$\a_R(x_0) = \a(x_0/R)/R$, $\int d x_0\, \a(x_0) = \tilde{\a}(0) =
1$, cannot exist as an operator in $\H$, only its  exponential
$V^C(f_R \a_R)$, formally  $\exp{ [i \, C_0(f_R \a_R)]}$, may be
defined. Furthermore, as shown by the following Proposition, such
exponentials have vanishing expectation on Gauss invariant states,
i.e. their restriction to the physical states vanishes.
\begin{Proposition} The operators $V^C( \l f_R \a_R)$,
formally the exponentials $\exp{ [i \l\,C_0(f_R \a_R)]}$ of  the
regularized space integrals $C_0(f_R \a_R)$ of $C_0(x)$, and
therefore assumed to transform under gauge transformations as such
exponentials, cannot be weakly continuous in $\l$ and therefore
the field $C_0(f_R \a_R)$ cannot be defined.

Furthermore,   the $V^C(f_R \a_R)$ satisfy, for all Gauss
invariant vectors $\Psi, \,\Phi$, \be{ (\Psi,\, V^C(f_R
\a_R)\,\Phi) = 0.}\ee
\end{Proposition}
\Pf\,\, In fact, if $C_0(f)$, $f \in \D(\Rbf^4)$,  exists, by
using the Gauss gauge invariance of the vacuum state $\omega$, the
vanishing of $\omega(A_k)$ by rotational invariance and
eq.\,(3.4), one has $$\omega(C_0(f)) = \omega(V(\U^\l))\, C_0(f)
V(\U^\l)^{-1}) =$$ $$= \omega(C_0(f)) + \int d^3 x
f(\x,t)\,n_{\U^\l}(\x).$$ Since for any $f$ there is at least one
$\U^\l(\x)$ such that the last term on the right hand side does
not vanish, one gets a contradiction. Thus, only the exponential
of $C_0(f)$ can be defined.

\noindent Moreover, given  $f_R$ one can find a small gauge
transformation $\U(\x)$, with $$\U(\x) = \U_1(\x)\,
\U_2(\x),\,\,\,\,\,n_{\U_1} + n_{\U_2} = 0,\,\,\,\,\,\,n_{\U_1}
\neq 0, $$ $$f_R \,\U_2 = \id, \,\,\,\,\,\,f_R\, \U_1 = \U_1. $$
Then, $\partial_i f_R \,\partial_j \U = 0$ and the second term on
the right hand side of eq.\,(3.4) vanishes; furthermore $\int d^3
x\, n_\U(\x) f_R(\x) = n_{\U_1}$. Hence, one has $$ (\Psi,\,
V^C(f_R \a_R)\,\Phi) = (\Psi,\,V(\U)\, V^C(f_R
\a_R)\,V(\U)^{-1}\,\Phi) =$$ $$\hspace{14mm}=\,\, e^{ i
n_{\U_1}}\,(\Psi,\, V^C(f_R \a_R)\,\Phi)$$ and eq.\,(3.4) follows.

\vspace{1mm} In conclusion, one cannot directly exploit the
non-invariance of $C_0(x)$ for proving a non-trivial action of the
large gauge transformations in $\H'$; it is essential to take into
account the non regularity of $\exp{i \,C_0(f_R \a_R})$, its non
observability and the non-existence  of the limit $R \ra \infty$.

%%%%%%%%%%%%%%%%%%%%%%%%%%%%%%%%%%%%%%%%%%%
%%%%%%%%%%%%%%%%%%%%%%%%%%%%%%%%%%%%%%%%%%

%%%%%%%%%%%%%%%%%%%%%%%%%%%%%%%%%%%%%%%
%%%%%%%%%%%%%%%%%%%%%%%%%%%%%%%%%%%%%%%
%\newpage
\section{Chiral symmetry and  solution of the $U(1)$ problem}

The situation changes substantially  in the presence of massless
fer\-mions, since the role of the topological current is taken by
a {\em conserved} current; hence, there is a symmetry associated
to it and the crucial point is its  relations with the
implementers of the large gauge transformations.

In this case, the Lagrangean, eq.\,(2.1) gets modified by the
addition of the (gauge invariant) fermion Lagrangean and the Gauss
operators become $$G_a = (\Dbf \cdot \Ebf)_a - j^a_0,
\,\,\,\,\,j_\mu^a = i g \bar{\psi} \g_\mu t^a \psi.$$
 The time independent gauge transformations of the fermion fields in the
 fundamental representation of the gauge group  are
$$\a_\U(\psi(x)) = \U(\x) \psi(x).$$

At the classical level, the Lagrangean is invariant under the
one-parameter group of  {\em chiral transformations} $\b^\l$, $\l
\in \Rbf$, $$\b^\l(\psi) = e^{ \l \g_5} \,\psi,
\,\,\,\,\,\b^\l\,(\bar{\psi}) = \bar{\psi}\,\, e^{ \l
\g_5},\,\,\,\,\,\g_5^* = - \g_5,\,\,\,\,\,\,\b^\l({\bf A}) =
\Abf.$$ Correspondingly, there is a conserved current $j_\mu^5 = i
g \bar{\psi}  \g^5 \g_\mu \psi$, the gauge invariant fermion axial
current.

In the quantum case, a gauge invariant point splitting
regularization is needed for the definition of $j^5_\mu$ and this
inevitably leads to an {\em anomaly},  $$\partial^\mu j_\mu^5 = -
2
\partial^\mu C_\mu =  - 2 \P.$$ The conserved axial current is now the gauge
dependent current $$J_\mu^5(x) = j_\mu^5(x) + 2 C_\mu,$$ its
conservation being equivalent to the anomaly equation for
$j^5_\mu$.
\goodbreak

For the discussion of the Weyl quantization, we take as local
exponential field algebra $\F_W$ the algebra generated by the
operators $W(\fbf)$,   by the gauge invariant bilinear functions
of the fermion fields and  by the unitary operators $V^5(f)$, $f
\in \S(\Rbf^4)$, formally the exponential of $J^5_0(f)$.

As  shown by Bardeen~\cite{B} on the basis of perturbative
renormalization in local gauges, the above (time independent)
chiral transformations of the fermion fields are generated by the
quantum field operator $J_\mu^5(x) $ and not by the gauge
invariant non conserved current $j_\mu^5$ ; the continuity
equation of $J_\mu^5 $ plays a crucial role in Bardeen analysis.

This justifies  our  assumption that the  exponential field
algebra contains the (formally defined) exponential of the smeared
field $J_\mu^5$ and in particular of the regularized integral of
the charge density $J_0^5(f_R \a_R)$; the corresponding
one-parameter group of unitary operators is denoted by $V^5_R(\l)
= V^5(\l\,f_R \a_R)$.

With the same motivations given before, eqs.\,(2.7)  are assumed
to hold together with the following transformation law of
$V_R^5(\l)$ under gauge transformations: for $R$ large enough so
that $f_R(\x) = 1 $ on the localization region of $\U_n(\x)$, one
has \be{ \a_{\U_n}(V^5_R(\l)) = e^{i\,2 n\,\l} V_R^5(\l).}\ee This
relation formally reflects the transformation properties of
$C_0(f)$ and the gauge invariance of $j_0^5$.
 By
the proof of Proposition 3.3, $V_R^5(\l)$ is not weakly continuous
in $\l$ and its formal generator $J^5_0(f_R \a_R)$ does not exist.
However, $V_R^5(\l)$ act as local implementers of $\b^\l$: in
fact, since $J_\mu^5$ is conserved, most of the standard wisdom is
available~\cite{MS2} and one has \be{ \limR V^5_R(\l)\, F\,
V^5_R(-\l) = \b^\l(F), \, \, \,\,\,\,\, \forall F \in \F_W.}\ee

 It is important to stress
that, thanks to locality, the above limit is reached for finite
values of $R$, and that it preserves locality and gauge
invariance. Thus, contrary to what is  stated in the literature,
the presence of the chiral anomaly does not prevent the chiral
symmetry from being a well defined time independent automorphism
of the field algebra $\F_W$ of its gauge invariant (observable)
subalgebra.

The loss of chiral symmetry is therefore a genuine phenomenon of
spontaneous symmetry breaking and the confrontation with the
Goldstone theorem becomes a crucial issue, the so-called ${\bf
U(1)}$ {\bf problem}.

%%%%%%%%%%%%%%%%%%%%%%%%%%%%%%%%%%%%%%%%
%%%%%%%%%%%%%%%%%%%%%%%%%%%%%%%%%%%%%%%%%%%%%%%%

%%%%%%%%%%%%%%%%%%%%%%%%%%%%%%%%%%%%%

%\section{Solution of the $U(1)$ problem}

The absence of parity doublets requires that the chiral symmetry
be  broken and the $U(1)$ problem amounts to explaining the
absence of the corresponding Goldstone massless bosons.

As discussed above, one of the basic assumptions of the Goldstone
theorem, namely the existence of an automorphism of the algebra of
observables, which commutes with space and time translations is
satisfied.

The second crucial property, needed the proof of the theorem, is
the local generation of the symmetry by a conserved current at
least in expectations on the vacuum state, i.e. \be{ \frac{d}{d
\l} < \b^\l(A) >_{\l = 0} = < \delta^5(A) > = i \limR <
[\,J_0^5(f_R \a_R),\,A\,]
>. }\ee Since the chiral automorphism $\b^\l$ is $C^\infty$ in
$\l$ its generator $\delta^5$ is well defined, but the problem is
its relation with the  formal generator  of the unitary
one-parameter group defined by the  $V_R^5(\l)$.

As a matter of fact, even if $\b^\l$ can be described by the
action of the local operators $V^5_R(\l)$, eq.\,(4.2), the
non-regularity of the one-parameter unitary group $V^5_R(\l)$,
prevents the existence of the corresponding generator $J_0^5(f_R
\a_R)$, so that one cannot write the symmetry breaking Ward
identities and obtain the Goldstone energy-momentum spectrum.
%\goodbreak

Quite generally, one has
\begin{Proposition} If $\omega$ is a gauge invariant vacuum state
and $A$ is an observable symmetry breaking order parameter, i.e.
$$\omega(\b^\l(A)) \neq \omega(A) \neq 0,$$ (the standard
candidate being $\bar{\psi} \psi$) then the expectations
$\omega(J_0^5(f_R \a_R)\,A )$ cannot be defined and eq.\,(4.3)
does not hold.
\end{Proposition}
\Pf\,\,In fact, otherwise,  for $R$ sufficiently large, one would
have \be{\omega(J_0^5(f_R \a_R)\,A ) = \omega(\a_{\U_n}(J_0^5(f_R
\a_R)\,A)) = \omega(J_0^5(f_R \a_R)\,A) + 2 n\,\, \omega(A),}\ee
i.e. a contradiction.

\vspace{2mm} Clearly, by  Proposition 3.1, the above Proposition
applies to the Gauss invariant state $\Psio$. The  impossibility
of writing  expectations involving $J_\mu^5$ on a gauge invariant
vacuum state, solves the problems raised by R.J. Crewther in his
analysis of chiral Ward identities.~\cite{Cr}

It is worthwhile to remark that, for  the evasion of the Goldstone
theorem discussed above, the occurrence of the so-called chiral
anomaly (which is present also in the abelian case) is not enough;
the crucial ingredient  is eq.\,(4.1), which directly implies the
non-regularity of the unitary operators $V^5_R(\l)$ and the
non-existence of the  local charges $J_0^5(f_R \a_R)$ in
expectations on a gauge invariant vacuum state.
%\newpage

%%%%%%%%%%%%%%%%%%%%%%%%%%%%%%%%%%%%%%%%%%%%%%%

%%%%%%%%%%%%%%%%%%%%%%%%%%%%%%%%%%%%%%%%%%%%%%%%%%%
\section{Topology, chiral symmetry breaking  and vacuum structure}
\def \vn {V(\U_n)}
\def \tvn  {\vn}
\def \svn {S(\U_n)\,}
\def \ein {e^{ i 2 n \l}}
The implications of the non-trivial  topology of the gauge group
on the breaking of the chiral symmetry crucially involve the
chiral transformations of the implementers of  gauge group. Such
chiral transformations do not follow merely from eq.\,(4.1),
because eq.\,(4.2) applies to the elements of the local field
algebra and its extension to  implementers, even if they are
strong limits of elements of $\F_W$, is not uniquely defined. This
problem does not arise if there are implementers $\tvn$ belonging
to the local field algebra.

Under this assumption, we shall first  prove that the chiral
symmetry is broken in any irreducible (or factorial)
representation of the field algebra, as well as in any factorial
representation of the observable algebra  in the Gauss invariant
subspace $\H'$ (we recall that a representation  is factorial if
the center of the strong closure consists of  multiples of the
identity there).

We shall later show that the standard vacuum structure is obtained
by analyzing the decomposition of $\H'$ into factorial
representation of $\A$ ($\theta$ sectors), in the case of a
reducible representation of the field algebra defined by a chiral
invariant vacuum; a crucial ingredient for the derivation are the
non-trivial chiral transformations of the implementers $\tvn$,
which uniquely follow from eqs.\,(4.1), (4.2), since they belong
to the field algebra.

\vspace{2mm} The local structure of the gauge transformations
imply that the implementers $V(\U)$ commute with the fields
localized in spacetime regions $\O$ disjoint from the spacetime
localization region $C_\U$ of $\U$; thus the $V(\U)$ are local
with respect to the field algebra, with localization region of
$V(\U)$ given by $C_\U$. This locality property partly motivate
the  assumption of existence of {\em local} implementers, in the
following sense.

As it is standard, in the following the local field algebras
$\F_W(\O)$ and their gauge invariant (``observable'') subalgebras
$\A(\O)$ are taken as strongly closed; $\F_W$ and $\A$ will denote
their unions over $\O$. For any bounded region $\O$, the local
algebra $\F_W(\O)$ may be identified with  the strong closure of
the polynomial algebra generated by the  field exponentials
$W(\fbf)$, by the gauge invariant bilinear functions of the
fermion fields and by the unitary operators $V^5(f)$; as it is
standard, the center of $\F_W(\O)$ is assumed to be trivial.

 Then, we assume that if supp $\U$ is localized in $ \O$, the
gauge transformation $\a_{\U}$ may be implemented by a unitary
operator $V(\U) \in \F_W(\O)$.~\cite{BDLR} Since the center of
$\F_W(\O)$ is trivial, such a local implementer is unique; in
particular, for Gauss trasformations $\U^\l$ it reduces to the
previously introduced Gauss operators.

Then, we have

\begin{Proposition} The local implementers  $V(\U)$  of the  gauge
transformations  are of the form $V(\U) \,F \,\Psio = \a_{\U}(F)\,
C_{\U}\,\Psio$, where $C_\U $ commutes with $\F_W$ and belongs to
the strong closure of $\F_W$. $C_{\U_n}$ depends only on $n$ and
on the Gauss invariant vectors one has \be{ \vn \,\Psi = C_n \Psi,
\,\,\,\,\,\,\forall \Psi \in \H'.}\ee The algebra generated by the
$C_n$ is abelian and one has \be{ C_n \,C_m = C_{n + m}.}\ee
Furthermore, one has
\newline i) for any $\O$,  there exist local operators $C_n(\O)$ belonging to
the  local observable algebra $\A(\O)$, satisfying $ [\,C_n(\O),
\, \A\,] = 0$   and \be{ C_n(\O)\,C_m(\O) = C_{n + m
}(\O),\,\,\,\,\,\, C_n(\O) \,\Psi = C_n \,\Psi, \,\,\,\,\,\forall
\Psi \in \H', \,\,\,\forall \O, }\ee
\newline ii) $ s-\lim_{\O \ra \Rbf^4}
\,C_n(\O) = C_n\,P_0, \,\,\,\,\,\,P_0 \H = \H'.$
\end{Proposition}
\def \FW {\F_W}
\def \an {\a_{\U_n}}
 \Pf \, In fact, if $F \in \F_W(\O)$, given $\a_{\U_n}$, the
transformation $\a_{\U_n}\,\a_{\U^a_n}^{-1}$, with $\U_n^a(\x)
\eqq \U(\x - \abf)$, is a Gauss gauge transformation and therefore
it is implemented by a product $V^{(a)}$ of  Gauss operators
$V^{(a)}(\U^i)$, which leave the vacuum vector invariant. Then,
the following strong limit exists and defines an operator
$$S(\U_n) \eqq s-\lim_{|\abf| \ra \infty} V^{(a)}, $$ which
satisfies $$\svn F\, \svn^{-1} = \a_{\U_n}(F), \,\,\,\,\forall F
\in \F_W, \,\,\,\,\,\svn \Psio = \Psio.$$ Clearly, $C_{\U_n} \eqq
\vn \svn^{-1} $ commutes with $\F_W$, belongs to the strong
closure of $\FW$ and gives $\vn F \Psio = \an(F)\,C_{\U_n} \Psio$.

\noindent Since $C_{\U_n}$ commutes with $\F_W$, they are
completely characterized by their action on $\Psio$, where they
only depend on the topological number $n$, commute with space time
translations and satisfy eq.\,(5.2), because $V(\U'_n)^{-1}\,\vn$,
and in particular $V(\U^a_n)^{-1}\,\vn$, are products of Gauss
operators, which leave $\Psio$ invariant.

 \noindent Any open set contains an open cylinder $\O$, with base
$\O_0$, $\O = \O_0 \times (t_1, \,t_2)$,  we denote by $\H'(\O) $
the subspace of vectors invariant under all the Gauss operators
$V(\U^\l)$, with supp $\U^\l \subset \O_0$ and by $P(\O)$ the
corresponding projector, which belongs to (the strongly closed)
$\F_W(\O)$. Then, as in the proof of Proposition 3.2, one has that
$\forall \,\U$,  $ V(\U^\l)\, V(\U) = V(\U)\, V((\U')^\l),$ with
supp $(\U')^\l = $ supp \,$\U^\l$; therefore
\be{[\,V(\U),\,P(\O)\,] = 0.}\ee The operators $C_n(\O) \eqq
V(\U_n)\, P(\O)$, with supp \,$\U_n \subset \O_0$, only depend on
the topological number $n$, since gauge transformation $\U_n,
\U'_n$ of the same homotopic class supported in $\O_0$ differ by a
product $\U_G$ of Gauss transformations with the same support and
$V(\U_G)\, P(\O) = P(\O)$. This implies the first of eqs.\,(5.3).

\noindent Furthermore, since  $\forall \, \U$, $V(\U)\,V(\U_n) =
V(\U'_n)\,V(\U),$ supp \,$\U'_n \subset \O_0$, by eq.\,(5.4), one
has $$V(\U) \,C_n(\O) =
 C_n(\O)\,V(\U),$$ i.e. $C_n(\O)$ are gauge invariant
and therefore belong to $\A(\O)$. Clearly, by construction
$C_n(\O)$ commutes with $\A$.

\noindent Since $P(\O) \Psi = \Psi$, $\forall \,\Psi \in \H'$, and
$\cap_{\O} \H'(\O) = \H'$, the remaining statements  follow.

\vspace{2mm} The unitary operators $C_n$ are related to the
topology of the gauge group, but they {\em do not} implement the
(large) gauge transformations on the field algebra;  they provide
a unitary  representation of the gauge group modulo the subgroup
of Gauss transformations, through operators in the commutant,
actually in the center, of $\F_W$.

 The important feature of the operators $C_n(\O)$ is that
of providing a  representation of the  group of gauge
transformations localized in $\O$, modulo Gauss gauge
transformations, through local operators which belong to the
center $\Z(\O)$ of the observable algebra localized in $\O$. Thus,
{\em the non-trivial topology of the gauge group is reflected by a
non-trivial center of the local observable algebras} $\A(\O)$.

\def \tvn {\vn}
The  above structure implies  that the non-trivial topology of the
gauge group and eq.\,(4.1) force the breaking of chiral symmetry.
\goodbreak
\begin{Proposition}  Chiral symmetry $\b^\l$ is  broken in
any irreducible (or factorial) representation of the field algebra
$\F_W$, defined by a Gauss invariant vacuum, as well as  in any
factorial (sub-)representation of the local observable algebra
$\A$ in the Gauss invariant subspace $\H'$.
\end{Proposition}
\Pf \,If the chiral symmetry $\b^\l$ is unbroken in $\H$, then
there is a one-parameter group of unitary operators $U^5(\l)$, $\l
\in \Rbf$ satisfying \be{ \b^\l(F) =
U^5(\l)\,F\,U^5(-\l),\,\,\,\,\, \,\forall F \in
\F_W,\,\,\,\,\,\,\,U^5(\l)\,\Psio = \Psio.}\ee Such an action of
$U^5(\l)$  provides the unique strongly continuous extension of
$\b^\l$ to $\bar{\F}_W$. Therefore, since the operators $\svn$ are
strong limits of Gauss operators, which are invariant under chiral
transformations, by eqs.\,(4.1), (4.2), one has $$\b^\l(S(\U_n)) =
S(\U_n).$$ On the other hand, by the localization of the $\vn$,
eq.\,(4.2) applies and then eq.\,(4.1) (for $R$ sufficiently
large) gives \be{U^5(\l)\,\tvn U^5(-\l)= \b^\l(\tvn) = e^{i 2 n
\l}\, \tvn;}\ee therefore \be{\b^\l(C_n) = U^5(\l) C_n U^5(-\l) =
\ein C_n.}\ee This is inconsistent with an irreducible (or
factorial) representation of $\F_W$, where the $C_n$ are multiples
if the identity.

\noindent Similarly, as a consequence of eqs.\,(4.1), (4.2), one
has $$\b^\l(P_0) = U^5(\l) \,P_0\,U^5(-\l) = P_0$$ and
\be{\b^\l(C_n P_0) = \ein \,C_n\,P_0.}\ee Since $C_n\ P_0$ is the
limit of elements of the (strongly closed) local observable
algebras $\A(\O)$, it belongs to the center of the representation
of $\A$ in $\H'$; this implies the instability of any  factorial
subrepresentation of $\A$, under $\b^\l$.

\vspace{2mm} The link between the non-trivial topology of the
gauge group and the labeling of the factorial representations of
the local observable algebra ($\theta$ sectors) is clearly
displayed in a (reducible) representation of the field algebra
defined  by a chirally invariant vacuum state. Such an invariance
arises in an analysis  based on the functional integral
formulation and semiclassical considerations,~\cite{CDG} ~\cite{C}
as well as in rigorous treatments of soluble models ({\em in
primis} the Schwinger model~\cite{CF}~\cite{MPS}~\cite{S}). In
general,  one obtains chirally invariant correlation functions by
using chirally invariant  boundary conditions in the functional
integral in finite volume.~\cite{LMS2}~\cite{BMS}

We shall therefore consider the case in which   chiral symmetry is
implemented in $\H$ by a one-parameter group of unitary operators
$U^5(\l)$, i.e. eqs.\,(5.5) hold.
\begin{Proposition} Under the above assumptions
the factorial subrepresentations, $\pi_{\theta}$, of $\A$ in $\H'$
are labeled by an angle $\theta$ ($\theta$ sectors):
$\pi_\theta(C_n) = e^{i\,2 n \theta} \id$, $ \theta \in [0, \pi)$,
(the corresponding groundstates are  called $\theta$ vacua).
\end{Proposition}
\Pf \, By eq.\,(5.2), $C_n = C_1^n$ and by eq.\,(5.7) the spectrum
of $C_1$ is $\s(C_1) = \{ e^{i 2 \,\theta}; \theta \in [\,0, \,
\pi)\} $.This is also the spectrum of the operator $C_1 P_0$ in
$\H'$, by eq.\,(5.8).

\noindent The Hilbert space $\H'$ has a central decomposition over
the spectrum of $C_1 P_0$ in $\H'$. Thus, one has $$\H' =
\int_{\theta \in [0, \, \pi)} d\theta\, \H_{\theta}, \,\,\,\,\,
\,C_n \H_{\theta} = e^{i 2 n \theta }\,\H_{\theta},$$ since the
spectral measure can be taken invariant under translations by
eq.\,(5.7) and by the chiral invariance of the vacuum. By
eq.\,(5.7),  $U^5(\a)$ intertwines between the sectors \be{
U^5(\a)\,\H_\theta = \H_{\theta +  \a}}\ee and satisfies $U^5(\pi)
= (-1)^F$, with $F$ the fermion number ($= 0$, by our definition
of $\F_W$). Since the chiral symmetry commutes with time
translations, the spectrum of the Hamiltonian is the same in all
$\theta $ sectors and ``all the $\theta$ vacua have the same
energy''. The same decomposition applies to the representation of
$\F_W$ in $\H$.

 Such a picture is exactly the same as in the
quantum mechanical model of QCD structures discussed in
Ref.~\cite{LMS2}

\vspace{2mm} Equation (5.6) provides a correct derivation of the
equation \be{ [\,V(\U_n),\, Q^5\,] =  2\,n,}\ee which is at the
basis of most of the standard discussions of chiral symmetry
breaking in QCD.  The standard   derivation  assumes that the
gauge transformations of the axial charge density extend to its
space integral, giving the transformation properties of the chiral
charge $Q^5$. Our derivation of the relation between chiral
symmetry and large gauge transformations, eq.\,(5.6), does neither
require the (usually assumed) convergence of the space integral of
$J_0^5$ to $Q^5$, nor that of its exponential, which are
incompatible with Proposition 3.3.

It is worthwhile to stress that the breaking of chiral  symmetry
is governed by  quite a different mechanism with respect to the
Goldstone  or the Higgs mechanism. In all  the three cases the
symmmetry commutes with spacetime translations. However, in the
Goldstone case, the symmetry breaking order parameter, typically
an observable operator, has strong enough localization properties
(preserved under time evolution) and its transformations under the
symmetry are generated by a local conserved current. In the Higgs
case, in positive gauges like e.g. the Coulomb gauge, the symmetry
breaking order parameter is not an observable and it has a
non-local time evolution, so that the (time independent) symmetry
is not generated at all times by the associated conserved local
Noether current.

In the axial $U(1)$  case of QCD, contrary to statements appeared
in the literature, the chiral transformations define a time
independent symmetry of the observables. The Goldstone theorem,
i.e. the presence of associated massless Goldstone bosons, is
evaded by the impossibility of writing the corresponding symmetry
breaking Ward identities, since the associated conserved Noether
current does not exist, only its exponentials do (non-regular
representation of the field algebra).

Actually, the chiral symmetry cannot be locally generated  by
unitary operators in any factorial representation of the
observable algebra, because the local observable algebras have a
center which is not left pointwise invariant under the chiral
symmetry.

In conclusion, the non-regular Weyl quantization provides a
strategy for putting the derivation of the vacuum structure and
the chiral symmetry breaking in the temporal gauge of QCD in a
more acceptable and convincing mathematical setting.
%%%%%%%%%%%%%%%%%%%%%%%%%%%%%%%%%%%%

%%%%%%%%%%%%%%%%%%%%%%%%%%%%%%%%%%%%%%%%%%%%%%%%%%%%
%%%%%%%%%%%%%%%%%%%%%%%%%%%%%%%%%%%%%%%%%%%%%%%%%%%%%%

%%%%%%%%%%%%%%%%%%%%%%%%%%%%%%%%%%%%%%%%%%%%%%%%%%%%%%

%%%%%%%%%%%%%%%%%%%%%%%%%%%%%%%%%%%%%%%%%%%%%%%
%\newpage
\section{Regular temporal gauge}

As discussed in the abelian case, one may look for an alternative
realization of the temporal gauge, by weakening the condition of
Gauss gauge invariance of the vacuum, so that the corresponding
correlation functions of gauge dependent fields and not only those
of their exponentials may be defined.

To be more precise, as before, one introduces a local field
algebra $\F$, generated by $\Abf(\fbf)$, $f^i_a \in \S(\Rbf^4)$,
by the fermion fields, by  their gauge invariant  bilinears, by
the axial current $J_\mu^5$, and by local operators $V(\U)$, which
implement the time independent gauge transformations $\a_\U$,
eq.\,(2.7), represent the the group $\G$ and  satisfy
\be{\a_{\U_n}(J_0(f_R \a_R)) = J_0(f_R \a_R) + 2\, n.}\ee We
denote by $\A$ the gauge invariant (observable) subalgebra of $\F$
and by $V_G$ a generic monomial of the Gauss operators $V(\U^\l)$.

A {\bf regular  quantization of the temporal gauge} is defined by
a (linear hermitian) vacuum functional $\omega$ on $\F$, which is
invariant under space-time translations and rotations and such
that its restriction to the observable algebra $\A$ satisfies
positivity, Lorentz invariance and the relativistic spectral
condition.

>From a constructive point of view, such a realization of the
temporal gauge may be related to a functional integral
quantization with a functional measure given by the Lagrangean of
eq.\,(2.1) with the addition of the fermionic part (see Section
4). The invariance of the Lagrangean with respect to the residual
gauge group after the gauge fixing ${\bf A}^0 = 0$, does not imply
the corresponding residual gauge invariance of the correlation
functions of $\F$, as discussed in the abelian case,~\cite{LMS}
~\cite{LMS3} since an infrared regularization is needed which
breaks the residual  gauge invariance. Therefore, the Gauss
constraint does not hold anymore.

The correlation functions of $\F$ given by  an $\omega$ with the
above properties define a vector space $\D = \F \Psio$, with
$\Psio$ the vector representing $\omega$, and an inner product on
it $<\,.\,,\,.
>$, which is assumed to be left invariant by the operators
$V(\U)$.

 It is further assumed that
$\omega$   satisfies the following {\bf weak Gauss invariance}:
$$\omega(A \,V_G) = \omega(A), \,\,\,\,\,\,\forall A \in \A,
\,\,\,\,\,\,\,\forall V_G,$$ equivalently  \be{< A\,\Psio,
\,V_G\,\Psio > = < A\,\Psio, \,\Psio
>, \,\,\,\,\,\,\forall A \in \A, \,\,\,\,\,\forall V_G.}\ee It
follows that the vectors of the the subspace $\D_0'\eqq \A\,\Psio$
are weakly Gauss invariant in the sense of  eq.\,(6.2) and
furthermore the space time translations $U(a)$ leave $\D_0'$
invariant. Thus, $\omega$ defines a vacuum representations of $\A$
in which the Gauss law holds.

The  weak form of Gauss gauge invariance
 of the vacuum functional
allows for the existence of the fields of $\F$ as operators on
$\D$, but the inner product cannot be semidefinite on $\D$ (by the
argument of Proposition 2.1). The subspace of vectors $\Psi \in
\D_0'$ with null inner product, $< \Psi, \, \Psi >\, = 0$, is
denoted by $\D_0''$.

Now, there is a substantial difference  in the realization of the
chiral symmetry, with respect to the representation defined by a
Gauss invariant vacuum. Thanks to the weak form of the Gauss gauge
invariance, the (smeared) conserved current $J_\mu^5$ may be
defined as an operator in $\D = \F \Psio$ and the standard wisdom
applies; in particular, for the infinitesimal variation $\d^5 F$
of the fields under chiral transformations, following Bardeen, one
has \be{\d^5 A = i \limR [\, J_0^5(f_R \a_R), \,A\,],
\,\,\,\,\forall A \in \A. }\ee In general, the representation
$\pi^{(0)}$ of the observable algebra defined by the vacuum vector
$\Psio$ may not be  irreducible  and therefore in order to discuss
the breaking of the chiral symmetry one must decompose it into
irreducible representations. Even if $\omega(\d^5 A) = 0$, a
symmetry breaking order parameter may appear in the irreducible
components of $\pi^{(0)}$.  Furthermore, such a decomposition of
the vacuum functional on $\A$ does not {\em a priori} extend to a
decomposition of the vacuum expectations $<\Psio, \,J_0^5(f_R
\a_R)\, A\,\Psio >$, since $J_0^5(f_R \a_R)$ is not gauge
invariant. Thus, one of the basic assumptions of the Goldstone
theorem may fail and chiral symmetry breaking may not be
accompanied by massless Goldstone bosons.

More definite statements can be made under the following
reasonable assumption, hereafter referred to as the {\em existence
of local implementers of the gauge transformations}:

\noindent i) the subspace $\D'$ generated by the vectors $ V(\U)
\D_0'$, with $ \U$ running over $\G$, satisfies the weak Gauss
constraint and semi-definiteness of the inner product,

\noindent ii) if supp $\U \subseteq \O$, then, $V(\U)$ can be
obtained as a ``weak'' limit of polynomials $F_n$ of $A_i^a$ and
$\psi$   localized in $\O$, in the following sense \be{< \Psi,
V(\U) \Phi
> = \lim_{n \ra \infty} < \Psi, F_n \,\Phi >,
\,\,\,\,\,\,\forall \Psi , \Phi \in \D.}\ee

Property i) is supported by the fact that the states defined by
the vectors $V(\U) A \Psio$, $A \in \A$, are weakly Gauss
invariant and positive; in fact $\forall A, B, C \in \A$,  $$<
A\,V(\U)\,B \Psio, \, V_G \,V(\U) C \Psio
> = < \a_{\U} (A)\,B \Psio, \, V'_G C \Psio > = $$ $$= <
A\,B\,\Psio, \,C \Psio > = < A\,V(\U)\,\Psio, V(\U)\,C \,\Psio
>.$$
The stability under $V(\U)$ of a weakly Gauss
 invariant subspace, which includes $\D_0'$, is automatically
satisfied if such a subspace may be selected by a gauge covariant
subsidiary condition. Weak Gauss invariance of $\D'$ is also
implied by the following stronger form of the weak Gauss
invariance of the vacuum functional \be{\omega(A
\,V(\U^\l)\,V(\U)) = \omega (A\,V(\U)), \,\,\,\,\,\forall A \in
\A, \,\,\,\,\,\,\forall \U^\l, \U.}\ee Property ii) is  supported
by the localization of the gauge transformations so that the
$V(\U)$ are  local relative to the field algebra, with
localization region given by the support of the corresponding
gauge transformation.

The fields $F$ which leave $\D'$ invariant also leave the subspace
$\D''$ of null vectors of $\D'$ invariant and therefore define
unique gauge invariant operators $\hat{F}$ in the ``physical''
quotient space $\D_{phys} \eqq \D'/\D''$,  which is the  analog of
the Gauss invariant subspace $\H'$ of the non-regular realization
of the temporal  gauge. Thus, to all effects  such fields can be
considered as observable fields; in the following we shall take as
{\em observable algebra  localized in $\O$}, $\hat{\A}(\O)$, the
algebra of operators in $\D_{phys}$ generated by fields localized
in $\O$ which leave $\D'$ invariant and as observable algebra
$\hat{\A} \eqq \cup_\O \hat{\A}(\O)$.~\cite{SW} In particular, the
local operators $V(\U_n(\O))$ are weakly gauge invariant and
therefore they define unique operators $\hat{T}_{\U_n(\O)} \in
\hat{\A}(\O) $ in $\D_{phys}$.

By the same arguments discussed before, the $\hat{T}_{\U_n}$
depend only on $n$, are invariant under space time translations
and satisfy \be{\hat{T}_n \,\hat{T}_m = \hat{T}_{n +
m},\,\,\,\,\,\hat{T}_0 = \id.}\ee Moreover, since one may write
$\vn = V(\U^a_n) \,V_G$, for any local $F$, which leaves $\H'$
invariant, one has  \be{ \hat{T}_n \,\hat{F} = \hat{F} \,\hat{T}_n
.}\ee This implies that the $\hat{T}_n$ generated an abelian group
$\G_T$ and belong to the center $\Z(\O)$ of $\hat{\A}(\O)$,
$\forall \O$.

Furthermore, the local generation of the infinitesimal chiral
transformations, eq.\,(6.3), implies weak continuity of the
derivation $\d^5$ on  the local field algebras $\F(\O)$ and   by
property ii), the infinitesimal chiral transformations  of the
local implementers $V(\U_n)$ of the large gauge transformations
are determined  by eq.\,(6.1), i.e. $$< \D, \,\d^5(V(\U_n)\,\D
> = \lim_{m \ra \infty} < \D, \,\d^5(\,F_m)\,D > =$$
$$ \lim_{m \ra \infty} i < \D,\, [J_0^5(f_R \a_R),\,F_m)] \D
> = i < \D, [J_0^5(f_R \a_R),\, V(\U_n)]\, \D > =$$ \be{= i \, 2 n\, <
\D, \,V(\U_n)\,\D >,}\ee for $R$ sufficiently large so that
$f_R(\x) = 1$ on the localization region of $\U_n$. Thus, one has
\be{ \d^5(\hat{T}_n) = i \, 2 n \, \,\hat{T}_n,}\ee and if the
chiral symmetry is unitarily implemented in $\D_{phys}$ the
observable algebra (in $\D_{phys}$) has a non-trivial center $\Z$.

%\goodbreak
\begin{Proposition} Under the above general assumptions, one has

\noindent i) the non-trivial topology of the gauge group gives
rise to a center of the observable algebra (in the physical space
$\D_{phys}$), which not left pointwise invariant under the chiral
symmetry,

\noindent ii) the chiral symmetry is broken in any factorial
representation of the observable algebra,

\noindent iii) the decomposition of the physical Hilbert space
$\H_{phys} \eqq \overline{\D_{phys}}$ over the spectrum of
$\hat{T}_1$ defines representations of the observable algebra
labeled by an angle $\theta \in [0, \pi)$, giving rise to the
$\theta$ vacua structure,

\noindent iv) the expectations $\omega_\theta(J^5_0(f_R \a_R)\,
A)$, $A \in \A$, with $\omega_\theta$ invariant under gauge
transformations,
 cannot be defined and a crucial condition of the
Goldstone theorem fails.

\end{Proposition}
\Pf \,Most of the arguments are  essentially the same as in the
non-regular realization. In particular, an unbroken chiral
symmetry in a factorial representation of the algebra of
observable is incompatible with the non-trivial chiral
transformations of its center.

\noindent By eq.(6.9), the spectrum of $\hat{T}_1$ is $\{e^{i 2
\theta}, \,\theta \in [0, \pi)$, and, even if $J_0^5$ is well
defined as an operator in $\D$, the existence of the expectations
$\omega_\theta(J_0^5(f_R \a_R) \,A)$ would lead to the same
inconsistency as in eq.\,(4.4).

\sloppy
\section{The Schwinger model in the temporal gauge}\fussy
The general features discussed above are exactly reproduced by the
Schwinger model in the temporal gauge, usually regarded as a
prototype of the non-perturbative QCD structures; in particular,
the assumptions about the local implementers of the large gauge
transformations hold. \def \d1 {\partial_1} \def \do {\partial_0}

The bosonized Schwinger model in the temporal gauge is formally
described by the following Lagrangean density \be{ \L  = \ume (\do
\ph)^2 - \ume (\d1 \ph)^2 + \do \ph A_1 + \ume (\do A_1)^2, }\ee
where $\ph$ is the pseudoscalar field which bosonizes the fermion
bilinears and therefore is an angular variable, and $A_1$ is the
gauge vector potential.

The time evolution is formally determined by the following
canonical equations \be{ \pi = \do \ph + A_1, \,\,\,\,\do A_1 = E,
\,\,\,\,\,\do \ph = \Delta \ph, \,\,\,\,\,\do E = \do \ph.}\ee

\vspace{1mm}\noindent 1) {\em Representation by a Gauss invariant
vacuum}

\vspace{1mm} \noindent  As exponential field algebra, we take the
algebra generated by the unitary operators \be{ V_\ph(f), \,\,\int
dx \,f = n, \,\,\,\,V_A(h),
\,\,\,\,V_E(g),\,\,\,\,V_\pi(g),\,\,\,\,\,f, \,g,\,h,\,\in
\D(\Rbf), }\ee formally corresponding to the exponentials $e^{i
\ph(f)}$, $e^{ i A_1(h)}$, $e^{i E(g)}$, $ e^{i \pi(g)}$,
respectively, and satisfying the Weyl commutation relations, with
above restriction on $f$, required by the periodicity of $\ph$.

The time independent gauge transformations $$\a_\U(V_A(h)) = V_A(h
+\U \d1 \U^{-1}), \,\,\,\,\,\U(x_1) - \id \in \D(\Rbf),$$ \be{
\a_\U(V_\pi(g)) = V_\pi(g + \U \partial_1 \U),}\ee $\ph$ and $E$
being left invariant, are generated by the local operators
\be{V(\U) \eqq V_\ph(f)\,V_E(-f)  \eqq V(f), \,\,\,\,\,f = \U \d1
\U^{-1}, \,\,\,\,\int dx \,f = n .}\ee The gauge functions $f$
with $\int dx f = 0$, i.e. those of the form $f = \d1 g,$ $g \in
\D(\Rbf)$, define the Gauss transformations and those with $\int
dx f = n$, $n \neq 0$, define the large gauge transformations and
shall be labeled by the topological number $n$.
 Clearly, if $\U$
is localized in $\O$, equivalently supp\,$f \subseteq \O$, then
$V(f) \in \F_W(\O)$, so that  our assumption of local
implementability is verified. The dynamics is defined by
eqs.\,(7.2) and therefore $e^{ i \s(f)} \eqq e^{i(\ph - E)(f)}$ is
independent of time.

The chiral transformations $\b^\l$ are defined by
\be{\b^\l(V_{\ph}(f_n)) = e^{i 2 n \l} V_{\ph}(f_n), }\ee all the
other exponential fields being left invariant.  Thus, as argued in
general, the anomaly of the gauge invariant axial current $j_\mu^5
= \partial_\mu \ph$, $\partial^\mu j_\mu^5 = \eps_{\mu \,\nu}
\partial^\mu A^\nu$ does not prevent the chiral symmetry from
defining a one-parameter group of automorphisms of the
(exponential) field algebra  and of its gauge invariant subalgebra
$\A$, locally generated by the unitary operators $V_R^5(\l) \eqq
V_\pi(f_R \a_R)$.

The GNS representation of $\F_W$ by a Gauss invariant state
$\omega$ is characterized by a representative vector $\Psi_0$
which satisfies \be{V(\partial_1 g) \,\Psio = \Psio, \,\,\,\,\,\,g
\in \D(\Rbf) .}\ee The Gauss invariance of the vacuum vector is
independently required by the condition of positivity of the
energy, using  the positivity of the state $\omega$ and the
invariance under space translations. In fact, since $V(f)$
commutes with the Hamiltonian, one can take $\Psio$ as an
eigenstate of $V(f)$, i.e. $V(f) \Psio = e^{i \l(f)} \Psio$; then,
by introducing $$V(t) \eqq e^{i \a_t( \pi(g) - \Delta A(g))} =
V(0) e^{i t \Delta\s(g) + i t \int d x \,\Delta(1 - \Delta)
g^2},$$
\goodbreak \noindent one gets $$(V(0) \Psio,\, H\, V(0) \Psio = i
(d/d t) (V(0) \Psio, V(t) \Psio)_{t = 0} = $$ $$= - \l(\Delta g) -
\int d x \Delta (1 - \Delta) g^2.$$ Therefore, the positivity of
the energy $\forall g$ requires that the functional $\l(\Delta  g)
$ be $\leq 0$, $\forall g$, and therefore  $=0$, since it is
linear in $g$. On the other hand, since any $f$ can be decomposed
as the sum $h_1 \int d x f + h_2 \int d x \, x f, \, +  \Delta
h_3$, with $h_i \in \D(\Rbf)$, the invariance under space
translations requires $\l(h_2) = 0$ and one gets $\l(f) = 0$,
$\forall f = \d1 g$.

By the same argument of Proposition 2.1, the Gauss invariance of
the vacuum vector implies the vanishing of all the expectations
$\omega(F V_A(h))$, with $F$ any  element of the gauge invariant
subalgebra $\A$ of $\F_W$, unless $h = 0$. Hence, we are left with
the correlation functions of the gauge invariant fields.

Since  eqs.\,(7.2) imply $\square E + E = \Delta \s$, and $E$ is a
pseudoscalar field, the vacuum correlations functions of $E$ are
those of a free pseudoscalar field of mass $=1$. By the Gauss
invariance of the vacuum, the expectations $\omega(V(f_n)\,
\a_t(V_E(g))$ depend on $f_n$ only through the topological number
$n$ and therefore define operators $T_n$, which are invariant
under spacetime translations and satisfy $T_n \,T_m = T_{n+m}$.

The  residual arbitrariness is therefore that of the
representation of the abelian algebra $G_T$, generated by the
operators $T_n$ in the subspace $\A \Psio$. The $\theta$ vacua are
characterized by the expectations \be{ \omega_\theta(V(f_n)\,A) =
\omega_{\theta}(T_n A) = e^{i 2 n \theta} \omega_\theta(A),
\,\,\,\,\,\,\forall A \in \A.}\ee

On the other hand, the reducible representation defined by a
chirally invariant vacuum is characterized by the expectations
$$\omega(V(f_n)\,A) = \omega(T_n\, A) = \delta_{n\, 0}\,\omega(A),
\,\,\,\,\,\forall A \in \A, \,\,\,\b^\l(A) = A.$$

It is easy to check that all the general features of the QCD case
discussed in Sections 3-5, in particular the evasion of the
Goldstone theorem, the breaking of chiral symmetry in any
irreducible or factorial representation of the observable algebra,
as a consequence of the non-trivial topology of  the gauge group
and the $\theta$ vacua structure  are exactly reproduced.

%\newpage
\vspace{2mm} \noindent 2) {\em Regular representation}

\vspace{1mm} \noindent As local field algebra $\F$ we take the
canonical algebra generated by the fields $A_1(h)$, $E(g)$,
$V_{\ph}(f)$,  with $\int d x \,f = n$, $\do \ph(f)$ with  the
 equal time commutation relations $$[\,A_1(x_1,t), \,E(y_1,t)\,] =
i \delta(x_1 - y_1),$$ $$ [\,V_{\ph}(f), \, (\do \ph + A_1)(g)\,]
= i \int d x_1 \,f g \,V_{\ph}(f).$$

\def \d {\delta} The euclidean functional integral corresponding to the
Lagrangean of eq.\,(7.1) yields well defined correlation functions
of $E$ and $\partial_\mu \ph$ satisfying the (weak) Gauss law
constraint. \def \d {\delta} As before, the two-point function
$W_E(x)$ of $E$ is that of free pseudoscalar massive field. The
correlation functions of $e^{ i \ph}$ involve a zero mode $\ph_0$
and crucially depend on the boundary conditions (in finite
volume). Any boundary condition satisfying positivity yields the
(weak) Gauss law holds for the expectations of all the gauge
invariant variables. Periodic boundary conditions in finite volume
give chiral invariant correlation functions and therefore for any
polynomial function $\P$, $$< e^{i \ph_0}
 \P( \partial_1 \ph, E)> = \d_{n \,0} < \P(\partial_1 \ph, E)
 >,$$(for a general discussion of the role of the boundary
 conditions in QCD and in related models see Ref.~\cite{MS3}
 ~\cite{BMS}~\cite{LMS2}).

An infrared subtraction is needed for  defining the correlation
functions of $A_1$. The most general two-point function $W_A$ of
$A_1$ must satisfy
 $- (d^2/ d t^2)
W_A(x) = W_E(x)$ and therefore has the following form
($\omega(k_1) \eqq \sqrt{k_1^2 + 1}$) $$\frac{1}{2 \pi}\int d k_1
\,\omega(k_1)^{-3} \,e^{i (\omega(k_1)x_0 - k_1 x_1)} - \ume
i\,(\delta(x_1) - \ume e^{-|x_1|}) x_0\,$$ \be{ +B(x_1) x_0 +
C(x_1).}\ee Locality requires $B(x_1) = - B(-x_1)$, $C(x_1) =
C(-x_1)$. The term linear in $x_0$ violate positivity. As
discussed in the QED case~\cite{LMS} the function $C$ can be
removed by an operator time independent gauge transformation and
$B=0$ if $w$ is invariant under the $CP$ symmetry $\Gamma$:
$\Gamma(\ph(x_1)) = \ph(-x_1)$,  $\Gamma(A_1(x_1)) = A_1(-x_1)$.

The two-point function $< \partial_1 \ph(x_1, 0)\,A_1(y_1, 0) >$
parametrizes the infrared regularization of the functional
integral corresponding to the Lagrangean of eq.\,(7.1) and can be
taken to vanish. Then, all the two-point functions involving
$\partial_1 \ph$ and $A_1$ are determined. The corresponding
$n$-point-functions can be taken as factorized (Gaussian).

Then, we are left
 with the correlation functions involving the
zero mode $\ph_0$, equivalently the correlation functions
involving $e^{i \s(f_n)}$, $n \neq 0$.

 By the
weak Gauss invariance, the one-point function $ <e^{i \s(f_n)}>_0
\eqq s_n$ only depends on the topological number and $$s_n =
\bar{s}_{-n}, \,\,\,\,\,\,< e^{i \s(f_n)}\,e^{i \s(f_m)}
> = < e^{i \s(f_n + f_m)} > = s_{n + m}.$$ Semi-definiteness of
the subspace $\A \Psio = \D'$ implies that the sequence $\{s_n\}$
is of positive type and (since $e^{i \s(f_n)}$ commutes with $\A$)
the vacuum  functional on $\A$ has the decomposition \be{w(A) =
\int_0^\pi d \mu(\theta)\, \omega_\theta(A), \,\,\,\,\,\,\forall A
\in \A,}\ee $$\omega_\theta(e^{i \s(f_n)} \,A) = e^{i 2n
\theta}\,w(A), \,\,\,\,\forall A = \P(\partial_1 \ph, E).$$ The
representation of $\A$ is the same as in positive case; in fact,
it only depends on the weak Gauss invariance of the vacuum.

The (non-positive) extension to the gauge field algebra $\F$ is
given by the correlation  functions $< e^{i
\s(f_n)}\,A_1(z_1)...A_1(z_k) >, $ $z_i = (x_1^{(i)}, x_0^{(i)})$.
For simplicity, we consider the case of a chirally invariant
vacuum functional $w$. Then, all such correlation functions for $n
\neq 0$ vanish and $s_n = \d_{n 0}$, corresponding to $d
\mu(\theta) = d \theta/\pi$.

In agreement with the general analysis of Sections 3,4, the chiral
symmetry cannot be locally generated in the physical space
$\D_{phys} = \D'/ \D''$; in particular, the density of the axial
current $J_0^5 = \partial_0 \ph + A_1$, which generates the
symmetry on $\F$, cannot be defined there, by the argument of
Proposition 4.1. Thus, the breaking of the chiral symmetry in any
factorial representation of the observable algebra does not
require the existence of massless Goldstone bosons.

Since the correlation functions $< e^{i \s(h)} \P(\partial_1 \ph,
E, A_1) >$ factorize in terms of two-point functions, which
satisfy the cluster property,  the limits of the correlation
functions  of $e^{i \s(f_n^a)}$, $f^a_n(x_1) \eqq f_n(x_1 - a)$,
when $|a| \ra \infty$, exist and define the analogs  of the
operators $C_n$ of Section 5. Then, by the chiral invariance of
$w$, one has $w(C_n) = \delta_{n\, 0}$ and $w$ may be decomposed
as a direct integral of (indefinite) functionals $w_\theta$ on
$\F$, characterized by the expectations $w_\theta(C_n) = e^{i 2 n
\theta}$, which do not lead to a decomposition in which the $C_n$
are multiples of the identity. Clearly, $w_\theta$ coincides with
the $\theta$ vacuum on $\A$: $ w_\theta(A) = \omega_\theta(A)$,
$\forall A \in \A$, but represents a non-positive extension to
$\F$, which is not invariant under the gauge transformations.

The corresponding chiral symmetry breaking Ward identities
$$w_\theta([\,J^5_R,\,e^{i \s(f_n)}\,]) = i 2n \,w_\theta(e^{i
\s(f_n)}) = i 2n\,e^{i 2n \theta},\,\,\,\,\,\,\,\,J_R^5 \eqq
J_0^5(f_R \a_R),$$ involve correlation functions $< J_0^5(x)
\,e^{i \s(f_n)}
>$ which are independent of time, but the time independent
vectors, playing the role of the Goldstone bosons, are equivalent
to $e^{i 2 n \theta}\, \Psi_\theta$ in the physical space
$\H_\theta$ and therefore do not give rise to zero energy states
different from  the vacuum. Actually, the possibility of writing a
chiral symmetry breaking Ward identity in terms of a commutator
with a current operator in the physical space (with the vacuum
vector $\Psi_\theta$ in their domain) is excluded, since any such
a commutator with $e^{i \s(f_n)}$ has vanishing vacuum
expectation.

It easy to check that all the general features and assumptions of
the QCD case are realized.

\end{document}